\documentclass[aps,prl,twocolumn,groupedaddress,showpacs]{revtex4}

\usepackage{graphicx}

\begin{document}

\title{Distillation of Bose-Einstein condensates in a double-well potential}

\author{Y. Shin}
\author{M. Saba}
\author{A. Schirotzek}
\author{T. A. Pasquini}
\author{A. E. Leanhardt}
\author{D. E. Pritchard}
\author{W. Ketterle}

\homepage[URL: ]{http://cua.mit.edu/ketterle_group/}

\affiliation{Department of Physics, MIT-Harvard Center for
Ultracold Atoms, and Research Laboratory of Electronics,
Massachusetts Institute of Technology, Cambridge, Massachusetts,
02139}

\date{\today}

\begin{abstract}

Bose-Einstein condensates of sodium atoms, prepared in an optical
dipole trap, were distilled into a second empty dipole trap
adjacent to the first one. The distillation was driven by thermal
atoms spilling over the potential barrier separating the two wells
and then forming a new condensate.  This process serves as a model
system for metastability in condensates, provides a test for
quantum kinetic theories of condensate formation, and also
represents a novel technique for creating or replenishing
condensates in new locations.
\end{abstract}

\pacs{03.75.-b, 03.75.Lm, 64.60.My
}

\maketitle

The characteristic feature of Bose-Einstein condensation is the
accumulation of a macroscopic number of particles in the lowest
quantum state.  Condensate fragmentation, the macroscopic
occupation of two or more quantum states, is usually prevented by
interactions~\cite{Nozi95}, but may happen in spinor
condensates~\cite{Ho98,OM98}. However, multiple condensates may
exist in metastable situations. Let's assume that an equilibrium
condensate has formed in one quantum state, but now we modify the
system allowing for one even lower state.  How does the original
condensate realize that it is in the wrong state and eventually
migrate to the true ground state of the system?  What determines
the time scale for this equilibration process?  This is the
situation which we experimentally explore in this paper using a
double-well potential.

The process we study is relevant for at least four different
questions. (1) The description of the formation of the condensate
is a current theoretical frontier and requires finite-temperature
quantum kinetic theories.  There are still discrepancies between
theoretical predictions and experimental
results~\cite{GLB98,KDG02}. Our double-well system has the
advantage of being an almost closed system (little evaporation)
with well defined initial conditions and widely adjustable time
scales (through the height of the barrier). (2) Spinor
condensates show rich ground states and collective excitations
due to the multi-component order parameter~\cite{Ho98}. Several
groups have observed long-lived metastable
configurations~\cite{MSS99,SMC99,SEK03,CHB03} and speculated
about transport of atoms from one domain to another via the
thermal cloud~\cite{MSS99,SEK03}. The double-well potential
allows us to characterize such distillation processes in their
simplest realization.  (3) The incoherent transport observed here
in a double well-potential imposes stringent limitations on future
experiments aiming at the observation of coherent transport in
Josephson junctions~\cite{SFG97,GSF00,PS01}. (4) Our observation
of condensate growth in one potential well due to the addition of
thermal atoms realizes the key ideas of proposals on how to
achieve a continuous atom laser~\cite{HBG96} which is different
from the experiment where condensates were replenished with
transported condensates~\cite{CSL02}.

\begin{figure}
\begin{center}
\includegraphics{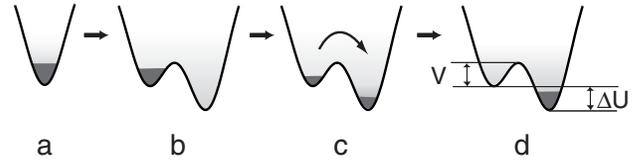}
\caption{Scheme for distillation of condensates in a double-well
potential. (a) Condensates are loaded into the left well. (b) A new
ground state is created by linearly ramping the trap depth of the
right well from zero to the final value. (c) Atoms transfer into the
right well via high-energy thermal atoms, and a new condensate starts
to form in the right well. (d) The whole system has equilibrated. $V$
denotes the height of the potential barrier between the two wells,
which is measured with respect to the bottom of the left well, and
$\Delta U$ the trap depth difference between the two wells.
\label{f:relaxation}}
\end{center}
\end{figure}

The scheme of the experiment is shown in Fig.\ref{f:relaxation}.
Bose-Einstein condensates in an optical dipole trap were prepared
in a metastable state by creating a second trap horizontally
adjacent to the first. Since the probability of quantum tunneling
through the barrier was extremely small~\cite{DGP99}, the coupling
between the two wells occurred only by the incoherent transfer of
high-energy thermal atoms over the potential barrier between the
two wells. The second trap was filled first by thermal atoms,
which then formed a new condensate. By monitoring the time
evolution of the double-well system we characterized how
differences in chemical potential and the height of the barrier
determined the dynamics.

Bose-Einstein condensates containing over $10^7$ $^{23}$Na atoms
were created in the $|F = 1,m_F = -1\rangle$ state in a magnetic
trap, captured in the focus of a 1064~nm optical tweezers laser
beam, and transferred into an separate ``science'' chamber as
described in Ref.~\cite{GCL02}. In the science chamber the
condensate was transferred from the optical tweezers into another
optical trap formed by a counter-propagating,
orthogonally-polarized 1064~nm laser beam.  As in
Ref.~\cite{SSP03}, the double-well potential was created by
passing a collimated laser beam through an acousto-optic modulator
(AOM) that was driven by two radio frequency (rf) signals. The
separation between the potential wells, $d$, was proportional to
the frequency difference, and the individual trap depth was
tailored by controlling the rf power at the two frequencies.
Typical parameters were an $1/e^2$ radius of each focused beam of
11.3~$\mu$m,  a single-well potential depth of $U = k_{B}
\times~2.4~\mu$K, where $k_{B}$ is the Boltzmann constant, and a
radial (axial) trap frequency, $f_r = 830$~Hz ($f_z = 12.4$~Hz).
As shown in Fig.\ref{f:relaxation}, condensates were initially
loaded into the left well with depth $U_L$ while the trap depth
of the right well, $U_R$, was maintained at zero. After holding
the condensates for 2~s to damp excitations which might have been
caused by the loading process, the temperature was $T_i = (180
\pm 90)$~nK, the number of condensed atoms $N_i = (1.1 \pm 0.1)
\times 10^6$ with a peak mean field energy of $\tilde{\mu}_0
\approx k_{B} \times~300$~nK, and the lifetime $\tau = (12.1 \pm
1.5)$~s.

The potential was transformed into a double-well potential by
linearly ramping the right well potential from zero to the final
value of $U_R$ over 500~ms while keeping $U_L$ constant.  This
time scale was chosen to be much longer than the radial trap
period of $\sim$1~ms to avoid excitations.  The resulting
double-well potential is characterized by the trap depth
difference between the two wells, $\Delta U = U_R - U_L$, and the
height of the potential barrier between the two wells, $V$, which
is measured with respect to the bottom of the left well, i.e. the
well initially full of atoms. The barrier height was set higher
than the peak atomic mean field energy of condensates so that
condensed atoms remained confined to the left well during the
transformation.

\begin{figure}
\begin{center}
\includegraphics{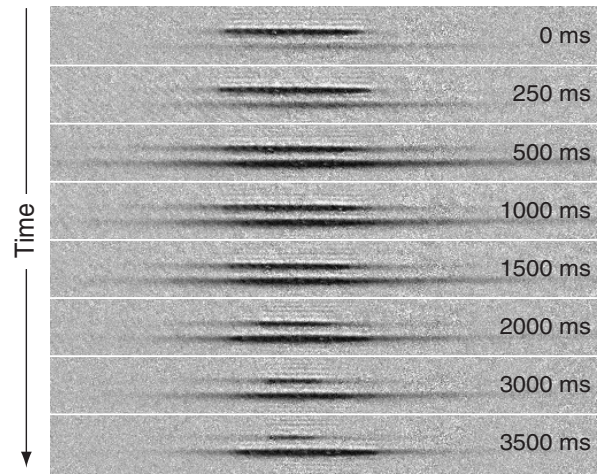}
\caption{Time evolution of atom clouds in a double-well potential.
The left (right) well appears as the top (bottom) atom cloud in
the images. A condensate was distilled from the left to the right
well. The absorption images were taken for various hold times
after creating the right well. The field of view of each
absorption image is 130~$\mu$m $\times$ 1160~$\mu$m. The trap
depths were $U_L = k_{B} \times~2.4~\mu$K (left well) and $U_R =
k_{B} \times~2.9~\mu$K (right well) with a potential barrier of
$V = k_{B} \times~510$~nK between them.  During the hold time,
the radial separation between the potential wells was $d =
15.9~\mu$m. \label{f:intrap}}
\end{center}
\end{figure}

The thermal relaxation process was observed by taking absorption
images of clouds confined in the double-well potential for various
hold times after turning on the right well. In order to fully
resolve the clouds in the two wells, their distance was increased
to $d=31.2~\mu$m just before taking absorption images. We assume
that this did not change either the number of atoms in each well
or the axial density distributions, since this additional
separation was done in 10~ms, which is much shorter than the
axial trap period of $\sim$100~ms, and the height of the
potential barrier exponentially increases when the two wells move
apart.

Fig.~\ref{f:intrap} shows the dynamical evolution for a situation
where the right well was much deeper than the left well.  In that
case,  condensates that initially existed only in the left well
were almost completely distilled within 3~s to form condensates
of comparable size in the right well.

The time evolution of the double-well system was characterized by
monitoring the number of condensed atoms and the temperature of
clouds in each well. These numbers were obtained by fitting
radially-integrated one-dimensional atomic density cross sections
to a bimodal distribution. The assumption of local equilibrium in
each well is justified by a short collision time $\tau_{\rm col}
\approx 1~$ms. For the condensate, we used a Thomas-Fermi
distribution, and for the thermal clouds, the fits to a
Bose-Einstein distribution were restricted only to the wings to
avoid the distortions due to the mean field repulsion of the
condensate~\cite{NS98}. The temperature turned out to be very
sensitive to the value of the chemical potential of the thermal
clouds. Assuming local equilibrium, we set the chemical potential
of the thermal clouds in each well equal to that of the
condensates in the same well.  In the absence of a condensate, the
chemical potential of the thermal cloud was determined by the fit
to a Bose-Einstein distribution

\begin{figure}
\begin{center}
\includegraphics{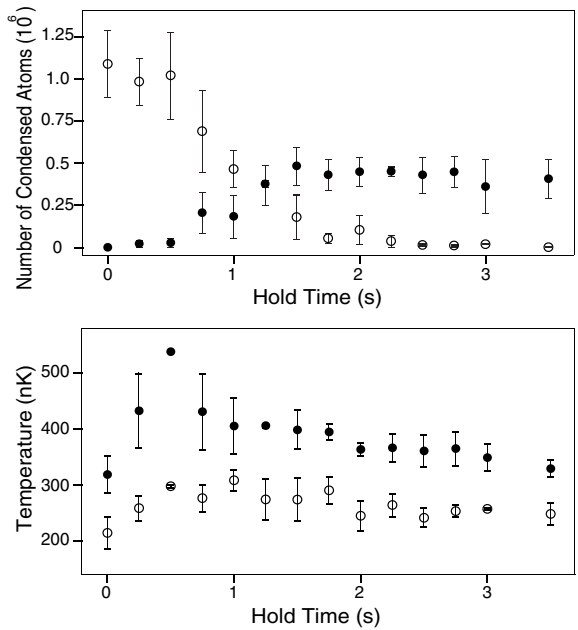}
\caption{Approach to thermal equilibrium in a double-well
potential. The temperature and the number of condensed atoms in
each well are shown as a function of hold time after creating the
right well. Open and solid circles represent atoms in the left
and the right well, respectively. Every data point is averaged
over three measurements, and the error bar shows $\pm$ one
standard deviation.  The experimental parameters are the same as
for the results shown in Fig.~\ref{f:intrap}. \label{f:growth}}
\end{center}
\end{figure}

Fig.~\ref{f:growth} displays the condensed atom number and
temperature for the images of Fig.~\ref{f:intrap}. Condensates
started to form in the right well after $(400 \pm 150)$~ms and
saturated within 2~s, resulting in $\sim 50\%$ of the condensate
being transferred. The final temperature in the right well was
$T_f \sim 350$~nK, which is $\sim 150$~nK higher than the initial
temperature $T_i$. This increase of temperature reflects the
energy gained by the atoms when they ``fall'' into the right
potential well which is deeper by $\Delta U = 480$~nK.  After
3.5~s, the total number of atoms of the whole system was $N_f =
(0.6 \pm 0.1) \times 10^6$, which is $15\%$ less than expected
for the measured lifetime of $\tau = 12.1$~s. Evaporative
cooling  due to finite trap depth may explain both the atom loss
and the fact that the temperature increase was much less than
$\Delta U$.

Even after 3.5~s hold time, full global equilibrium was not
reached. This can be seen in both the temperature and the
condensed atom numbers. As the chemical potential of condensates
in the right well was lower than the trap bottom of the left
well, there should not have been any condensate remaining in the
left well in global equilibrium. However, Fig.~\ref{f:intrap}
shows a small condensate of $\sim 10^3$ atoms in the left well
even after 3.5~s holding.  Furthermore, the temperature in the
left well was measured $\sim 100$~nK lower than in the right well.

On first sight, this slow approach towards equilibrium is
surprising. In evaporative cooling, one has very fast cooling for
a ratio of the height of the potential barrier to the temperature
of less than three~\cite{KvD96}, as was in our experiment. Note,
however, that in our trap geometry, the exchange of thermal atoms
is geometrically suppressed due to the small ``contact area''
between the two elongated cigar shaped clouds. Moreover, if the
transferred thermal atoms have high angular momentum, they have
poor collisional coupling to the cold trapped atoms like the Oort
cloud in magnetic traps~\cite{CEW99}. Indeed, the density of
thermal atoms with higher energy than the potential barrier in the
left well after 3.5~s holding is $\sim 3 \times 10^{11}$/cm$^3$,
and their collision time with the atoms confined in this well is
$(n \sigma v_{rel})^{-1} \approx 0.5 $~s.

Another quantity of interest in the condensate formation process
is the onset time of condensation, the hold time until a
condensate first appears~\cite{MSA98,GLB98,KDG02}. To avoid
ambiguities in fitting small condensates, we determined the onset
time in the right well by observing the appearance of
interference fringes when two condensates were released from the
double-well potential. For two pure condensates, the visibility
of the interference fringes is larger than $55\%$ as long as the
number ratio of the two condensates is larger than $\eta=$0.05.
Using the methods described in Ref.~\cite{SSP03}, we have
observed discernible interference fringes down to $\eta=0.08$,
corresponding to $\sim 8 \times 10^4$ condensed atoms in the
right well.

\begin{figure}
\begin{center}
\includegraphics{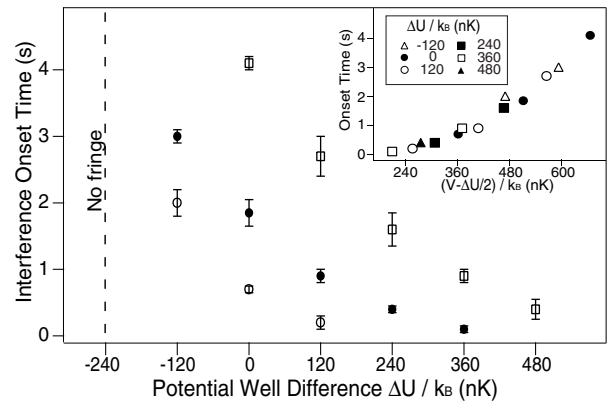}
\caption{Onset time of condensation. The onset time in the right
well was measured by observing the appearance of a matter wave
interference pattern when the condensates were released from the
double-well potential. The trap depth difference is defined as
$\Delta U = U_R - U_L$. $U_L$ was kept at $k_{B} \times 2.4~\mu$K
for all experiments. The separations of the two wells, $d$, were
14.3~$\mu$m (open circle), 15.1~$\mu$m (solid circle), and
15.9~$\mu$m (open square). Interference fringes were not observed
at $\Delta U = -k_{B} \times 240$~nK even after 20~s hold time.
The inset shows the same data plotted vs. $V-\Delta U/2$ where $V$
is the height of the potential barrier.\label{f:onset}}
\end{center}
\end{figure}

Onset times were measured as a function of $d$ and $\Delta U$
(Fig.~\ref{f:onset}).  The condensate formation is driven by the
potential well difference $\Delta U$, whereas the barrier of
height $V$ provides the `resistance' against equilibration, since
thermal atoms must have a kinetic energy larger than $V$ to
transfer from the left well to the right well. Phenomenologically
(see inset of Fig.~\ref{f:onset}) the condensate onset time
depends only on the combination $(V-\Delta U/2)$ with an almost
exponential dependence. $(V-\Delta U/2)$ can be considered as
$(V_{eff}-\Delta U)$, where $V_{eff}=[V+(V+\Delta U)]/2$ is the
average height of the barrier measured from each well.

In two limiting cases, no interference patterns were observed.
When the trap depth difference is larger than the peak atomic
mean field energy of condensates, i.e. $|\Delta
U|>\tilde{\mu}_0$, it is energetically favorable for condensates
to remain in the lowest well. We observed no interference pattern
when $\Delta U = -k_{B} \times 240$~nK even after 20~s hold
time.  The disappearance of interference fringes was observed
when $\Delta U \geq k_{B} \times 360$~nK due to complete
distillation of the condensates into the right well. In the limit
where the barrier height is smaller than the peak atomic mean
field energy of condensates, i.e. $V<\tilde{\mu}_0$, condensate
atoms can `spill' over the potential barrier. Indeed, we observed
that condensates appeared in the right well immediately for $V$
less than $\sim k_{B} \times 290$~nK, consistent with
$\tilde{\mu}_0 \sim k_{B} \times 300$~nK.

To observe quantum tunneling, the thermal relaxation time
$\tau_{th}(\propto exp[V/k_B T])$ should be longer than the
tunneling time $\tau_{tu}(\propto exp[\sqrt{V/m\hbar^2} w])$ where
$w$ is the thickness of the barrier. For a thick barrier like ours
($>5$~$\mu$m), the tunneling time is extremely long ($>10^5$~s)
and thermal relaxation is likely to dominate. A high and thin
barrier is necessary to observe tunneling and the related
Josephson effects.

In conclusion, we have created Bose-Einstein condensates in a
metastable state in a double-well potential and studied the
dynamical evolution. The observed distillation process is
important for equilibration in spinor condensates and for
replenishing condensates in continuous atom lasers.

This work was funded by ARO, NSF, ONR, and NASA. We thank K.~Xu
for a critical reading of the manuscript. M.S. acknowledges
additional support from the Swiss National Science Foundation.

\end{document}